\documentclass[apchd5,article]{achemso}

\usepackage{times}

\usepackage{graphicx}
\usepackage{textcomp}
\usepackage{color}
\usepackage{color,soul}
\usepackage{textgreek}

\title{Angle-resolved cathodoluminescence imaging polarimetry}
\author{Clara I. Osorio} 
\altaffiliation{These authors contributed equally to this work}
\author{Toon Coenen}
\altaffiliation{These authors contributed equally to this work}
\author{Benjamin Brenny}
\altaffiliation{These authors contributed equally to this work}
\author{Albert Polman}
\author{A. Femius Koenderink}
\email{f.koenderink@amolf.nl}
\affiliation{Center for Nanophotonics, FOM Institute AMOLF, Science Park 104, 1098 XG Amsterdam, The Netherlands}

\begin{document}

\begin{abstract}
Cathodoluminescence spectroscopy (CL) allows characterizing light emission in bulk and nanostructured materials and  is a key tool in fields ranging from materials science to nanophotonics.  Previously, CL measurements focused on the spectral content and angular distribution of emission, while the polarization was not fully determined. Here we demonstrate a technique to access the full polarization state of the cathodoluminescence emission, that is the Stokes parameters as a function of the emission angle. Using this technique, we measure the emission of metallic bullseye nanostructures and show that the handedness of the structure as well as nanoscale changes in excitation position induce large changes in polarization ellipticity and helicity. Furthermore, by exploiting the ability of polarimetry to distinguish polarized from unpolarized light, we quantify the contributions of different types of coherent and incoherent radiation to the emission of a gold surface, silicon and gallium arsenide bulk semiconductors. This technique paves the way for in-depth analysis of the emission mechanisms of nanostructured devices as well as macroscopic media.
\end{abstract}

 \maketitle

\section{Introduction}
Among many recent developments in microscopy, optical electron-beam spectroscopy techniques such as cathodoluminescence imaging (CL) have emerged as powerful probes to characterize materials and nanophotonic structures and devices. In CL, one collects light emitted in response to a beam of energetic  electrons ($0.1-30$ keV), for example in a scanning electron microscope (SEM). The time-varying evanescent electric field around the electron-beam interacts with polarizable matter creating coherent emission, such as transition radiation (TR) \cite{Adamo_PRL12,Bashevoy_OE07}. The spot size of the focused electron beam and the extent of the evanescent field about the electron trajectory define the interaction resolution to be below $\sim20$ nm, while the interaction time ($<1$ fs) determines the broadband character of the excitation.  Aside from coherent emission,  incoherent emission can also be generated both by the primary beam and by slower secondary electrons, which excite electronic transitions in matter \cite{Abajo_RMP07,Yacobi}. The relative importance of the coherent and incoherent contributions provides information about the material composition and electronic structure. Spectral analysis of the cathodoluminescence as a function of the electron beam position allows the local characterization of the structure and defects of semiconductors \cite{Edwards_SemicondSciTech2011,Sauer_PRL2000,Ton-That_PRB2012}, the functioning of nanophotonic devices \cite{Fontcuberta_PRB2009}, and to map the optical resonances of plasmonic and metamaterial structures \cite{Zhu_PRL10}. Recently developed techniques for 
detection of CL enable the identification of the band structure and Bloch modes of photonic crystals~\cite{Yamamoto_OE09,Yamamoto_OE11,Adamo_PRL12,Ma_JPC14,Sapienza_NM12}, the dispersion of surface plasmons~\cite{losstw,Bashevoy_OE07}, and the directivity and Purcell enhancement of plasmonic nano-antennas \cite{coenen_NL11,yamamoto_NL11}.

Besides frequency and linear momentum, the vectorial nature of light provides a third degree of freedom rich in information about the physics of light generation and scattering, encoded in the polarization of emitted light.  In materials characterization, for instance, polarization gives direct access to the local orientation of emission centers and  anisotropies in the host material. In nanophotonics, polarization  plays a fundamental role  (together with directionality) in determining the interaction between  emitters and nanostructures. Furthermore, it is increasingly recognized that mapping and controlling the polarization of light is key to harnessing the wide range of opportunities offered by metamaterials and metasurfaces. Recent breakthroughs in chirality-enhanced antennas \cite{Gorodetski_PRL13}, photonic topological insulators \cite{lu14}, and the photonic equivalent of the spin-Hall effect \cite{onoda04,yin13,li13,connor14}, indicate the emerging importance of  mapping  the full polarization properties of nanophotonic structures. Polarization measurements of CL emission, however, have been limited to fully polarized emission and in particular to linearly polarized signals~\cite{coenen_OE12,Coenen_NC14}.


In this letter we introduce a novel technique to access full polarization information in cathodoluminescence spectroscopy. Based on a polarization analysis method previously demonstrated in optical microscopes~\cite{fallet_MEMS11,arteaga_OE14,kruk_ACSP14, Osorio_SR15}, we integrate a rotating-plate polarimeter in the detection path of the angle-resolved CL setup. Using the Mueller matrix  formalism for the light collection system, we  determine the Stokes parameters for CL emission, that is, all parameters required to completely describe the polarization state of the light, which can be polarized, partially polarized or totally unpolarized.  We demonstrate the great potential of this new measurement technique by analyzing the angle-resolved polarization state of directional plasmonic bullseye and spiral antennas. Furthermore, and exploiting the unique capabilities of CL excitation, we measured the emission from metals and semiconductors. For these materials, we can separate coherent and incoherent emission mechanisms, with further applications in nanoscale materials science.


\section{CL Polarimetry}

\begin{figure*}[ht!]
\centering
\includegraphics[width=0.7\textwidth]{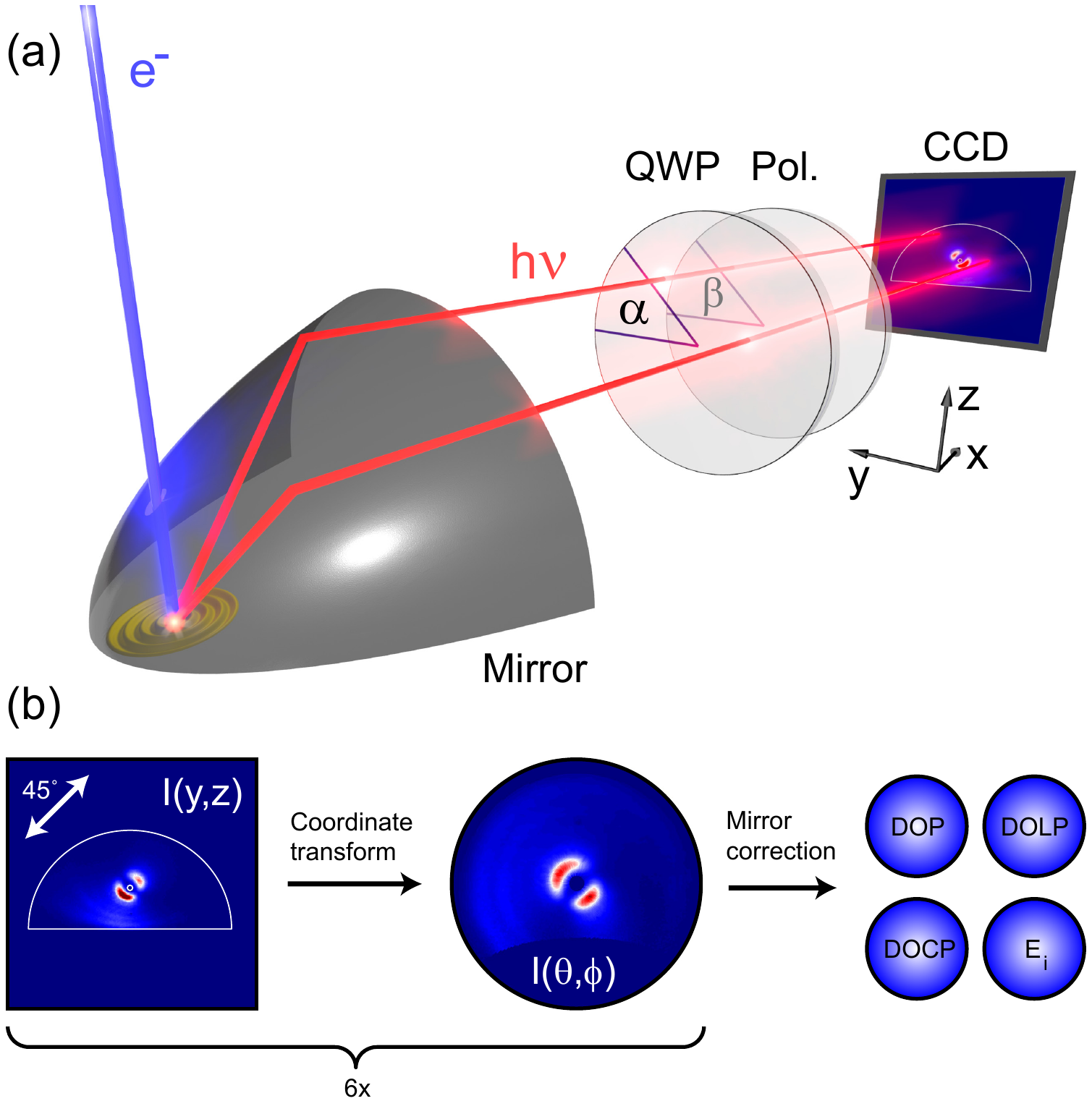}
\caption{(a) Schematic overview of the cathodoluminescence polarimetry setup. The structure is excited by the electron beam after which the resulting light emission is collected by a parabolic mirror. The light is directed towards a polarimeter composed of a QWP and linear polarizer set at angles $\alpha$ and $\beta$ respectively. The filtered beam profile is measured by the CCD camera. The CCD images shows  data corresponding to a measurement on a bullseye structure with $\alpha = \beta = 45^{\circ}$. For reference we also show the coordinate system that is used throughout the manuscript. (b) Six measurements with different settings of the polarimeter are required to retrieve the full angle-resolved polarization state of the collected light. In addition to a coordinate transformation, the effect of the mirror on the polarization is corrected for.  Using the retrieved Stokes parameters it is possible to determine any figure of merit for polarization including the total ($DOP$), linear ($DOLP$) and circular  ($DOCP$) degrees of polarization, as well as the electric field components $|E_i|$. } 
\label{Fig1}
\end{figure*}

 In our measurements, the $30$ keV electron beam from a scanning electron microscope (SEM) excites the sample. An aluminum paraboloid mirror collects and redirects  the resulting CL emission out of the SEM. The outcoming beam is focused onto a fiber-coupled spectrometer or projected onto a 2D CCD array~\cite{coenen_NL11,coenen_APL11,Sapienza_NM12}, as shown in Fig.~\ref{Fig1}(a). The wave-vector distribution of the CL emission can be retrieved from the CCD image, as every transverse point in the beam corresponds to a unique emission angle, in a procedure analogous to other Fourier imaging techniques~\cite{Lieb_JosaB04,Kosako_NP10,curto10,Aouani_NL11,Sersic_NJP11,Belacel_NL13}. 
 
Measuring  polarization for all emission angles of CL presents several challenges. First, it requires determining the relative phase difference between field components, a task not achievable with only linear polarizers as in  Ref.~\cite{coenen_OE12}. Second, the paraboloid mirror performs a non-trivial transformation on the signal as it propagates from the sample to the detector plane. The shape of the mirror introduces a rotation of the vector components of light  due to the coordinate transformation and, consequently, a change in the main polarization axes. In addition, the  angle and polarization-dependent Fresnel coefficients of the mirror modify the polarization of the light upon reflection ~\cite{Bruce_OPT06,Bruce_04}.  As a function of the angle of incidence, the mirror  partially polarizes unpolarized light and transforms linearly to elliptically polarized light.

To address these challenges, we included a rotating-plate polarimeter in the beam path of our CL system, composed of a quarter wave plate (QWP) and a linear polarizer \cite{Berry_ApplOp77,Born_Wolf,Chipman}. Figure~\ref{Fig1}(a) shows the polarizing elements in a schematic of the setup. Depending on their orientation, these two elements act either as a linear polarizer or as a right or left handed circular polarizer.  As  shown in Fig. \ref{Fig1}(b), we measure the intensities $I_j$ transmitted by six different settings of the polarimeter (horizontal, vertical,  $45^{\circ}$, $135^{\circ}$, right and left handed circular) in order to determine the Stokes parameters of the light:

\begin{eqnarray}
\label{stokes_eq}
S_0 &=&I_{H}+I_{V} \nonumber\\
S_1 &=& I_{H}-I_{V}\nonumber\\
S_2 &=& I_{45}-I_{135}\nonumber\\
S_3 &=&I_{RHC}-I_{LHC}.
\end{eqnarray}

\noindent  These four parameters are the most general representation of polarization and can be used to retrieve any polarization-related quantity \cite{Born_Wolf}. The raw polarization-filtered CCD images are projected onto [$\theta,\varphi$]-space as indicated in Fig.~\ref{Fig1}(b) using a ray-tracing analysis of the mirror, after which the Stokes parameters in the detection plane are determined. To transform these to Stokes parameters in the sample plane, we determine the Mueller matrix of the light collection system that accounts for the effects of the mirror on the polarization. In addition to the geometrical transformation, the analysis takes into account the Fresnel coefficients of the mirror for  $s$- and $p$- polarized light. Due to the 3D shape of the mirror, each element of the Mueller matrix is a function of the emission angle, i.e., there is a Mueller matrix for each emission angle. The supplementary information describes in more detail how the Mueller matrix was calculated and how we benchmark these calculations using fully polarized transition radiation (see Fig. S2).

\begin{figure*}[thb!]
\centering
\includegraphics[width=\textwidth]{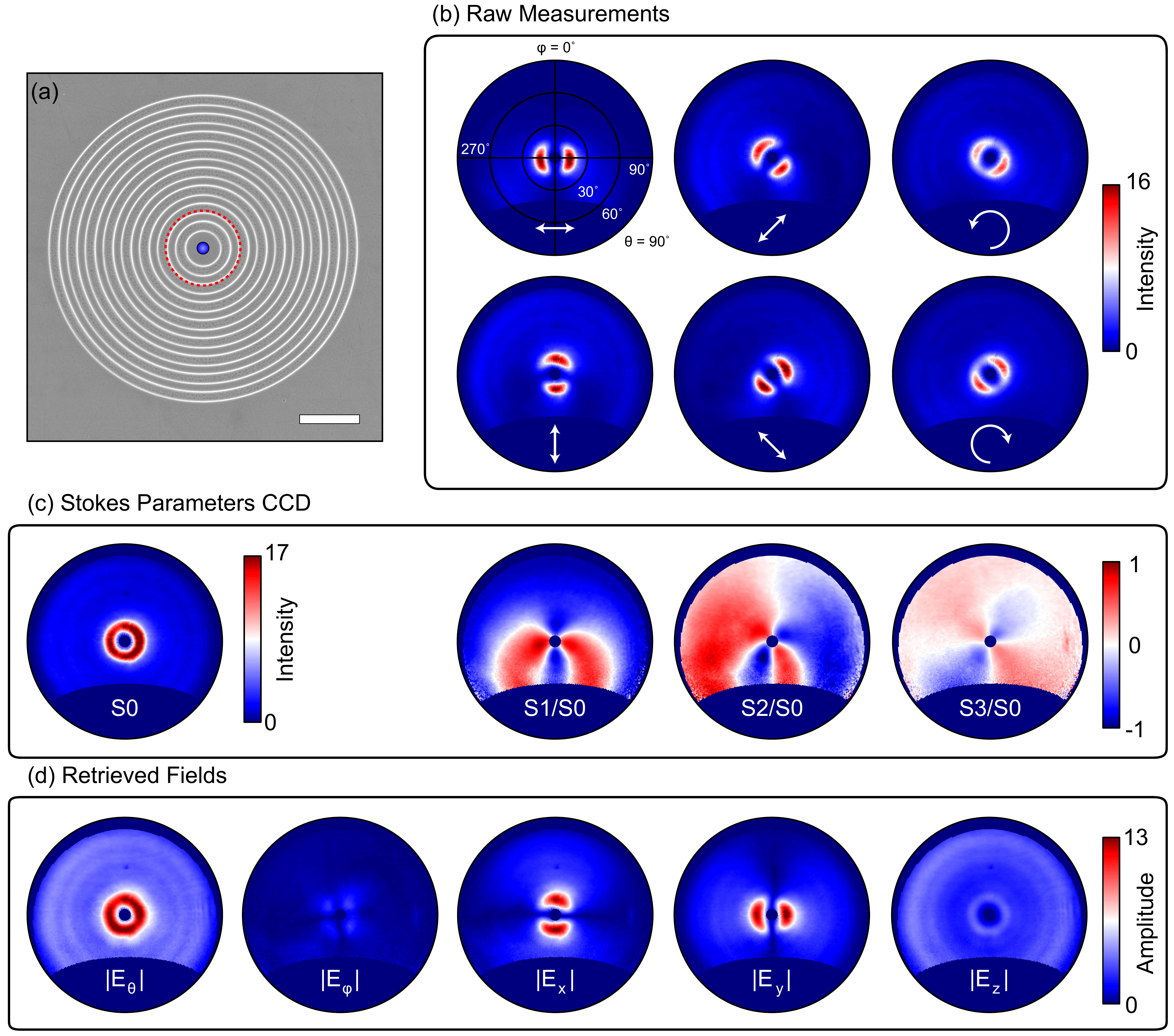}
\caption{(a) Scanning electron micrograph of a bullseye structure with $d = 600$ nm. The blue dot indicates the electron beam excitation position. The red dashed circle indicates which part of the bullseye is shown in detail in Fig.~\ref{Fig3}. The scale bar corresponds to $2$~\textmu m. (b) Polarization filtered angular CL patterns for different analyzer settings as indicated by the white arrows, measured at $\lambda_{0} = 750$ nm.  (c) Stokes parameters in the detection plane as a function of angle. The $S_{1}$, $S_{2}$, and $S_{3}$ patterns are normalized to $S_{0}$ to better show the overall polarization distribution. (d) Spherical and Cartesian field amplitude distributions as a function of angle, retrieved from the experimental data in (b). In all figures, the intensities are given in 10$^{5} ADU sr^{-1} s^{-1}$. Amplitudes are in units of $10^2 \sqrt{ADU sr^{-1} s^{-1}}$ (analog-to-digital units).}
\label{Fig2}
\end{figure*}

The Stokes parameters in the sample plane allow determining \emph{any}  figure of merit for polarization. Given that both incoherent and coherent radiation may be generated in CL, the degree of polarization ($DOP$), and the degrees of linear ($DOLP$) and circular polarization ($DOCP$) will be especially relevant. Defined as the ratio of polarized, linearly or circularly polarized light to total intensity, they are given by $DOP=\sqrt{S_1^2+S_2^2+S_3^2}/S_0$, $DOLP=\sqrt{S_1^2+S_2^2}/S_0$ and $DOCP=S_3/S_0$. Equivalently, the ratio of unpolarized light to total intensity is given by $1-DOP$ so  a $DOP$ smaller than $1$ corresponds to partially polarized light.

\section{CL Polarimetry on plasmonic structures}
\subsection{Bullseye antennas}
To demonstrate the full potential of angle-resolved CL polarimetry we investigate the emission of a plasmonic bullseye structure with a pitch $d = 600$ nm, milled into a single-crystal gold substrate. Bullseyes are well-known for their ability to strongly direct light scattered by nanoscale apertures~\cite{Lezec_Sc02}, generated by fluorescence~\cite{Jun_NC11,Aouani_NL11} or thermal emission~\cite{Norris_OE10}.  Figure~\ref{Fig2}(a) shows a scanning electron micrograph of the structure indicating the excitation position. The electron beam launches a circular surface plasmon polariton (SPP) wave which radiates outwards and scatters coherently from the grooves of the bullseye. The scattered fields interfere to give rise to directional emission.  In our measurements, the emission is spectrally filtered  by a $40$~nm bandwidth bandpass filter centered at $\lambda_{0} = 750$ nm (see Fig. S1 in the supplement for full spatial and spectral mapping).

Figures \ref{Fig2}(b-d) represent the main steps of our polarimetric analysis for CL. Figure \ref{Fig2}(b) shows the angular intensity patterns measured for the six settings of the polarimeter  (indicated by the arrows) after a coordinate transformation of the raw intensity data.  Figure~\ref{Fig2}(c) shows the Stokes parameters in the detection plane calculated using Eq.~\ref{stokes_eq} from the patterns in Fig.~\ref{Fig2}(b). The leftmost panel corresponds to the total intensity distribution, $S_0$. The bullseye emits in a narrow doughnut pattern without any azimuthal variations, consistent with the azimuthal symmetry of both excitation position and bullseye structure. The polar angle at which most of the CL is emitted, $\theta = 15^{\circ}$, corresponds to the grating equation $\theta = \sin^{-1}(k_{SPP}-m2\pi/d)/k_{0}$. In this spectral regime, the grating order $m=1$ is the only relevant order, $k_{0}=2\pi/\lambda_{0}$ and $k_{SPP}$ is the SPP wave-vector, calculated using the optical constants for gold from spectroscopic ellipsometry. The other panels in Fig.~\ref{Fig2}(c) show the Stokes parameters $S_{1}$, $S_{2}$, and $S_{3}$ in the detection plane ($yz$-plane in Fig.~\ref{Fig1}) normalized to $S_{0}$, such that it is possible to see polarization features outside the areas of very bright emission.

Next, we transform the data collected by the detector to the polarization state of the emitted light in the sample plane, by multiplying the Stokes parameters at the detection plane with the mirror's inverse Mueller matrix. Among the quantities that the Stokes parameters allow retrieving, here we will focus on the electric field components.  Figure \ref{Fig2}(d) shows the reconstructed spherical field vector amplitudes $|E_\phi|$ and $|E_\theta|$ that constitute the natural $s$- and $p$- polarization basis relevant to map the far-field generated by a localized radiating object. The figure shows that the $|E_{\theta}|$ distribution is strong and azimuthally symmetric while $|E_{\varphi}|$ is close to zero. Therefore, the measured emission of the bullseye is a narrow doughnut beam with a fully linear, radial polarization. Radial polarization is expected for the bulls-eye radiation as SPPs scatter out while maintaining their $p$-polarized character at the grooves. 

The polarization can alternatively be cast into Cartesian components. Figure \ref{Fig2}(d) shows the  double-lobe patterns of $|E_{x}|$ and $|E_{y}|$, which are rotated $90^{\circ}$ relative to each other. The $|E_{z}|$ component is azimuthally symmetric and shows several emission rings. The outer rings correspond to transition radiation (TR) from the excitation position, which is modulated to yield a fringe pattern due to interference with SPPs scattered off the bullseye grooves \cite{kuttge_PRB09}. Since the electric field must be transverse to the propagation direction, the $|E_{z}|$ component vanishes at near-normal angles and therefore the main SPP emission beam from the bullseye (the narrow ring) appears relatively weak in $|E_z|$. While the emission in the sample plane is completely linearly polarized, a nonzero circular polarized signal is measured in the detection plane ($S_3$ in Fig. \ref{Fig2}(c)), which indicates the effect of the mirror and the importance of using the Mueller matrix analysis to correct for it.   

\begin{figure*}[tbh!]
\centering
\includegraphics[width=0.8\textwidth]{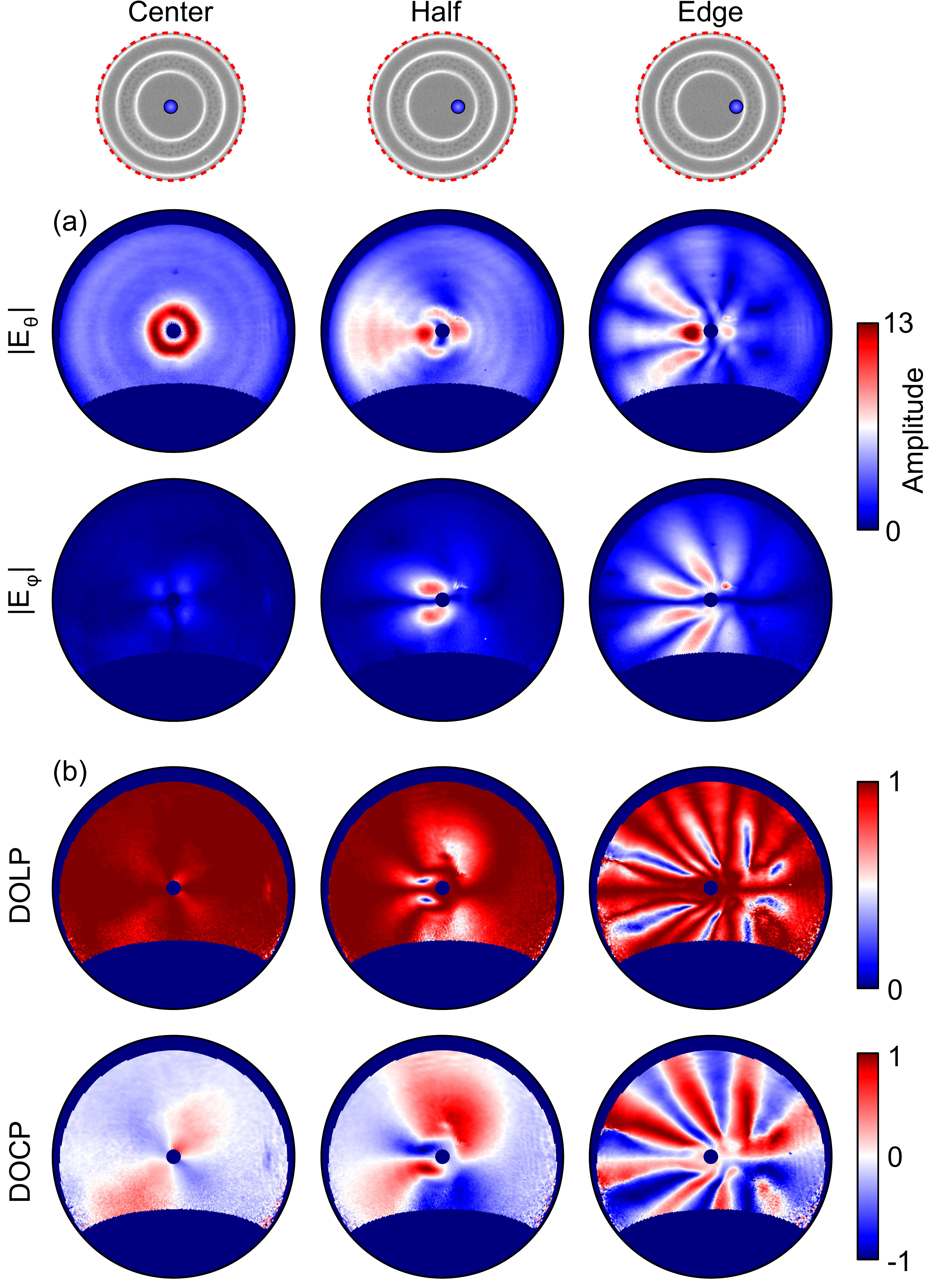}
\caption{(a) $|E_{\theta}|$ and $|E_{\varphi}|$ field amplitudes for central, halfway, and edge excitation on the bullseye plateau. The excitation positions are indicated as blue circles in the SEM micrographs on top, which show the area enclosed by the red dashed circle in Fig.~\ref{Fig2}(a). (b) Degree of linear ($DOLP$) and  circular ($DOCP$) polarization of the bullseye emission as a function of emission angle for the same excitation positions as in (a).} 
\label{Fig3}
\end{figure*}

\subsection{Non-symmetric geometries}
CL polarimetry is an unique tool to explore the relation between the symmetry of a system and its polarization response.  While the symmetry of a bullseye structure excited right at the center ensures the clear TM polarization emission shown in Ref. \label{Fig2}(d), this is not longer the case when launching an off-center circular SPP wave on the structure. Figure~\ref{Fig3} shows measurements for electron beam excitation in the center, halfway between the center and the edge, and at the edge of the central bullseye plateau, as indicated in the SEM micrographs on top of the figure. Figure \ref{Fig3}(a) shows $|E_{\theta}|$ and $|E_{\varphi}|$ for the three excitation positions. For off-center excitation, the zenithal field distribution $|E_{\theta}|$ is no longer symmetric, being stronger towards the left than towards the right of the image. This type of asymmetric beaming has also been observed in angular intensity measurements on asymmetric gratings~\cite{Jun_NC11}, spirals~\cite{Rui_SR13}, and asymmetrically excited (patch) antennas~\cite{Mohtashami_ACSP14,Coenen_NC14}. Besides the asymmetry, the off-center excitation also leads to a non-zero azimuthal field contribution, $|E_{\varphi}|$, which is similar in strength to the zenithal field contribution. The excitation position and the center of the bullseye defines a mirror symmetry that expresses itself as a nodal line for $|E_{\varphi}|$ at $\varphi = 90^{\circ}$ and $\varphi = 270^{\circ}$. At far off-center excitation, the azimuthal and zenithal field distributions are very rich in structure and for certain angular ranges the emission becomes elliptically or  circularly polarized.

The effect of off-center excitation is most clearly seen in Fig.~\ref{Fig3}(b), which shows the degree of linear ($DOLP$) and circular ($DOCP$)  polarization. Owing to the mirror symmetry of sample and excitation, the $DOCP$ remains close to zero along the axis of the electron beam displacement. Yet, away from this axis the emission becomes elliptical with opposite handedness on either side of the axis as dictated by mirror symmetry. For edge excitation, the complementary multi-lobe $|E_{\theta}|$ and $|E_{\varphi}|$ patterns lead to a rich behavior, where the emission changes from fully linear to almost fully circular polarization several times.

Rather than breaking symmetry by changing the excitation position, it is also possible to study scattering and emission by intrinsic asymmetry and handedness of structures such as Archimedean spirals. Spirals enjoy a growing interest since it was shown that they can enhance the extraordinary transmission of single nanoapertures for particular helicities \cite{Drezet_OE08}, and transfer polarization and orbital angular momentum to scattered photons \cite{Rui_SR13,Rui_OE12}. This can result in a polarization-dependent directional beaming \cite{Rui_OL11} and demonstrates strong photon spin-orbit coupling effects \cite{Gorodetski_PRL13}.

We fabricated Archimedean spiral gratings with clockwise (CW) and anti-clockwise (ACW) orientation, as shown in Fig.~\ref{Fig4}(a,b), and used CL polarimetry to study the effect of spiral asymmetry and handedness on the far-field polarization, again taking a pitch $d$ = 600 nm and $\lambda_{0}$ = 750 nm. We excite the spirals in their origin as indicated in Fig.~\ref{Fig4}. Figures~\ref{Fig4}(c,d) show the Cartesian components of the far-field emission of the spirals, which better reflect the handedness than the spherical fields. Since the groove pitch is the same for spirals and bullseyes, the angular spread of these patterns is similar, however they no longer have a minimum at the normal and the s-like shape in $|E_{x}|$ and $|E_{y}|$ clearly reflects the handedness of the spiral. 

\begin{figure*}[ht!]
\centering
\includegraphics[width=0.8\columnwidth]{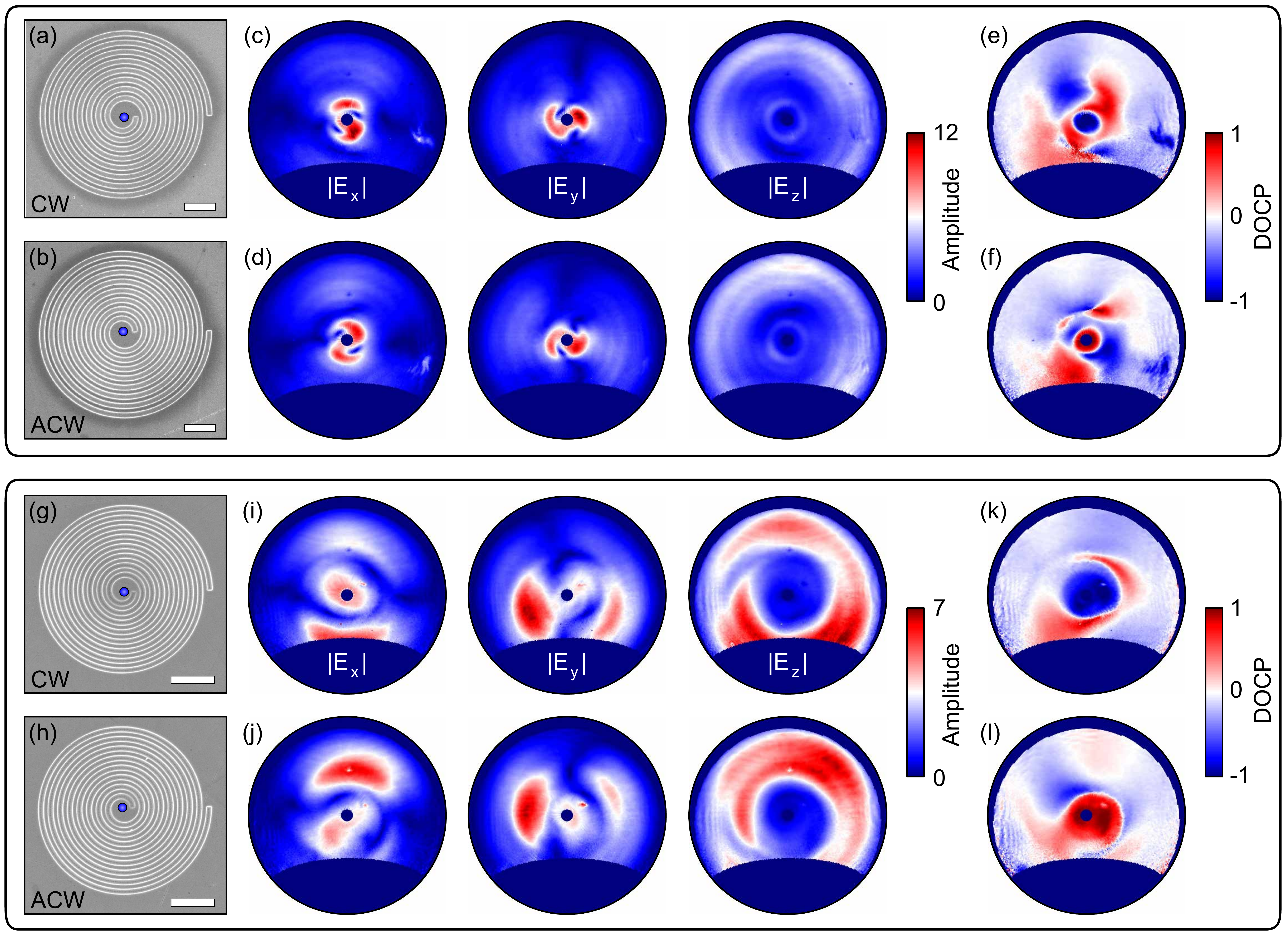}
\caption{Scanning electron micrographs of (a) a clockwise (CW) and (b) anti-clockwise  (ACW) Archimedean spiral grating with $d$ = 600 nm. Amplitude distributions of the Cartesian fields for (c) a CW spiral and (d) a ACW spiral. Degree of circular polarization ($DOCP$) for (e) a CW spiral and (f) an ACW spiral. Scanning electron micrographs of (g) a CW and (h) ACW spiral with $d$ = 440 nm. Amplitude distributions of the Cartesian fields for (i) a CW spiral and (j) a ACW spiral. Degree of circular polarization ($DOCP$) for (k) a CW spiral and (l) an ACW spiral. All measurements were performed at $\lambda_{0}$ = 750 nm. For reference we again indicate the electron beam excitation position with a blue dot. Scale bars in the electron micrographs correspond to 2 \textmu m.}
\label{FigS3}
\end{figure*}

In contrast with bullseyes excited at their center, the spirals can induce ellipticity in the polarization of the light even when excited in their origin, as shown in Fig.~\ref{Fig4}(e,f). This is particularly evident in the region of higher intensity in the vicinity of the normal, where the $DOCP$ is close to $\pm1$. Thereby, the spirals are highly directional sources of circularly polarized light. Mirrored spirals simply exhibit mirrored patterns (where the $y$-axis defines the mirror symmetry), conserving intensities and field strengths, while the sign of the $DOCP$ changes. This result indicates that swapping spiral handedness not only flips the helicity of the output field but, in addition, it mirrors the distribution of intensity over angle. For spirals with smaller pitch we find similar but even stronger effects of handedness, aided by the fact that their radiation pattern is more strongly off-normal (see Fig.~\ref{FigS3}(g-l)).
In that case  the $|E_{z}|$ distribution is also clearly chiral.

The data shown in this section proves that polarimetry analysis of CL, in combination with precise electron beam positioning,  provides direct insight into the complex emission behavior of nanophotonic structures. Measuring directionality and polarization of the emission from emitters coupled to single nanostructures is of paramount importance when designing and testing the performance of structures like optical antennas, plasmonic resonators, and metasurfaces.

\section{CL polarimetry applied to incoherent emitters}
\begin{figure*}[thb!]
\centering
\includegraphics[width=0.4\textwidth]{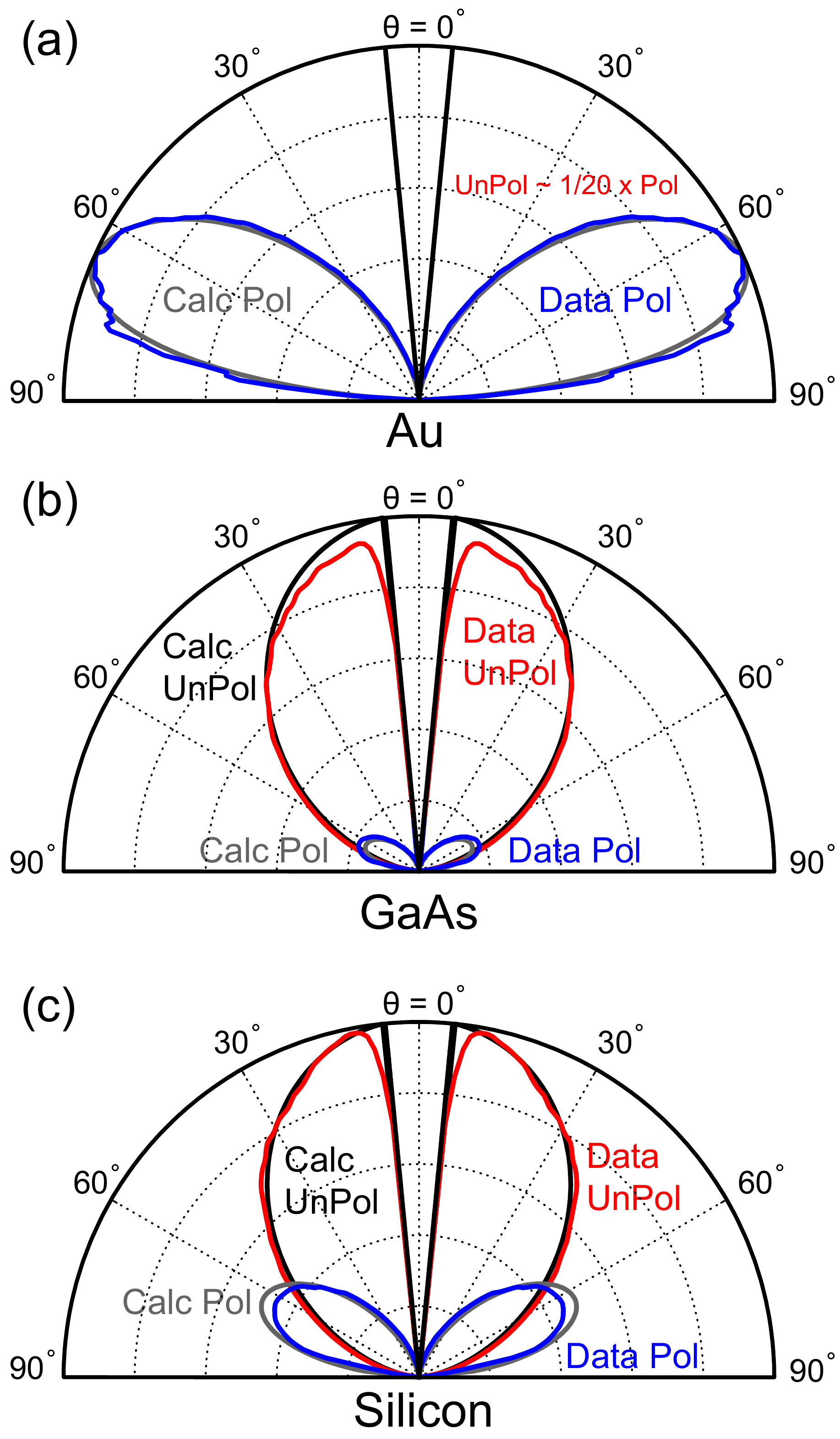}
\caption{Zenithal cross cuts comparing unpolarized and polarized emission for bulk single crystals of Au at $\lambda_{0} = 850$ nm (a), GaAs at $\lambda_{0} = 850$ nm (b)  and  Si at $\lambda_{0} = 650$ nm  (c). In all cases we compare the unpolarized (red) and polarized (blue) emission from measurements to the unpolarized (black) and polarized (grey) emission determined from calculations. The data is obtained by averaging over an azimuthal range $\varphi = 270^{\circ}$ -- $90^{\circ}$ to improve signal-to-noise ratio, and we scale the angular distributions by the overall emission intensity. For Au, the unpolarized emission component is so small that it is not visible in the plot.}
\label{Fig4}
\end{figure*}

 In addition to characterizing fully coherent radiation, CL polarimetry  allows us to determine whether the measured radiation contains an unpolarized contribution such as in the   case of incoherent luminescence from bulk or nanostructured materials. This is shown in Fig.~\ref{Fig4}, where we compare azimuthally averaged zenithal cross cuts of the polarized ($S_{0}\times DOP$) and unpolarized ($S_{0}\times(1-DOP)$) emission intensities for single-crystal Au, Si and GaAs and compare them to calculations.  

The emission from Au at $\lambda_0 =  850$ nm in Fig.~\ref{Fig4}(a) is expected for coherent TR (see also Fig. S2  in the supplement) and hence fully polarized. The data indeed shows excellent agreement with a calculated TR emission distribution. In the case of GaAs in Fig.~\ref{Fig4}(b), the emission is dominated by very bright incoherent radiative band-to-band recombination measured at $\lambda_{0} = 850$ nm. This luminescence is fully isotropic and unpolarized \emph{inside} the material, but large differences between $s$- and $p$- Fresnel transmission coefficients for the semiconductor-vacuum interface partially polarize the emission as seen in the data. Figure~\ref{Fig4}(b) shows that unpolarized light is indeed dominant. The weak polarized  emission has a very different angular emission distribution,  that agrees well with  Fresnel calculations (see Fig. S3 in the supplement for more information). Lastly, Si is a material that displays weak luminescence that it is comparable to TR~\cite{Brenny_JAP14}. Indeed, the polarized intensity for Si at $\lambda_{0} = 650$ nm shown in Fig.~\ref{Fig4}(c) constitutes $\sim 31.7\%$ of the total emission, which is much more significant than for GaAs, although unpolarized emission remains the dominating contribution.

These examples show that angle-resolved polarimetry measurements provide quantitative and precise information about the origin of emission of different materials. This technique enables the separation of polarized and unpolarized emission, and therefore it can be used to determine the different mechanisms that simultaneously contribute to cathodoluminescence (see Fig. S2 in the supplement for a quantitative analysis of TR). Moreover, for the polarized part of the emission we can map the electric field components and their relative phase. Since it does not require any prior knowledge of the sample (unlike the method described in Ref.~\cite{Brenny_JAP14}), this method is very general and can be applied to any (nanostructured) material.

\section{Conclusion}

We have demonstrated `angle-resolved cathodoluminescence imaging polarimetry' as a new microscopy tool to map the vectorial electromagnetic scattering properties of nanostructured and bulk  materials. We determine the complete polarization state of emitted light as a function of angle from six CL intensity measurements in the detection plane, in combination with a mathematical transformation that corrects for the polarizing effect of the CL mirror.  Due to the high resolution of the electron beam excitation, the wave-vector resolved polarization properties of locally excited plasmonic nano-antennas can be extracted with a spatial resolution for the excitation of $20$~nm.  The angle-resolved polarization measurements of the emission of bullseye and spiral nanoantennas demonstrate  how structural symmetry and handedness translate into the  helicity of emitted light. These results show that angle-resolved cathodoluminescence polarimetry can be extremely valuable for the development of metallic and dielectric antennas  for spin-resolved and chiral spectroscopy as well as for the study of photon spin Hall effects. 

Besides its relevance for nanophotonics, we demonstrate that our technique opens new perspectives for materials science not accessible with optical microscopes. Measuring the Stokes parameters generally enables the separation of incoherent and coherent CL generation, as we demonstrated for direct and indirect semiconductor materials. Our measurements on relatively simple samples of Au, GaAs and Si show the potential of the technique for the analysis of bulk materials which could be useful for many material inspection tasks. For optoeletronics, the nanoscale characterization of emission polarization from inorganic LEDs stacks, nanowires and quantum dots stands out in particular. The technique also introduces the possibility of locally studying material anisotropy, birefringence and optical activity.

\section{Methods}
\subsection{Sample fabrication} 
We fabricated bullseye and spiral structures by patterning a single-crystal Czochralski-grown Au $\langle100\rangle$ pellet which was mechanically polished to obtain a sub-$10$ nm RMS roughness. The patterning was done by using a $30$ keV Ga$^{+}$ ion beam in a FEI Helios NanoLab dual beam system at $9.7$ pA beam current and a dwell time of $10$ \textmu s per pixel. In the bullseye design the central plateau has a diameter of 1.2 \textmu m ($2$ times the pitch, $600$ nm) and the duty cycle of the circular grating consisting of $8$ grooves is $50\%$. The spiral design is based on an Archimedean spiral where the first half period of the spiral is omitted. For the spiral we show data both for $600$ and $440$~nm pitches. Both for the spirals and the bullseyes the groove depth was $\sim 110$ nm.
The measurements on silicon were performed on a polished p-type (boron doping level $10^{15}-10^{16}$ cm$^{-3}$) single-crystal $\langle 100 \rangle$ wafer. The measurements on GaAs were done on a polished single-crystal $\langle 100 \rangle$ wafer.

\subsection{Measurements} 
The measurements were performed in a FEI XL-30 SFEG ($30$~keV electron beam, $30$~nA current) equipped with a home-built CL system~\cite{coenen_NL11,coenen_APL11,Sapienza_NM12}. To obtain the polarization state of the emission, we perform a series of six measurement of the angular CL pattern using a 2D back-illuminated CCD array. Each measurement was taken in a different setting of the polarimeter, defined by a specific combination of QWP and polarizer angles. Handedness of circularly polarized light was defined from the point of the view of the source, following the IEEE standard. In this case, right-handed circularly polarized light rotates anti-clockwise and left-handed circularly polarized light rotates clockwise.  We use the known transition radiation pattern from an Au surface to calibrate the optical detection system.  A $40$~nm band pass color filter spectrally selected the measured emission. For the bullseye and spiral measurements we used $30$~s integration time, which is a good compromise between a small spatial drift of the electron beam and a good signal-to-noise in  CL. For the TR emission from single-crystal gold and the measurements on silicon we used $120$~s integration time since TR emission and luminescence are position independent and the measurement is not affected by spatial drift of the electron beam. For the measurements on GaAs we used a much lower current ($0.9$~nA) and integration time ($1$~s) due to the very bright band gap luminescence. For every setting of the polarimeter, we collected a dark reference measurement where we blank the electron beam (with the same integration time as the CL measurement), which was subtracted from the data in the post-processing stage.  Possible sources of errors on the measurements include e-beam drift (in the case of position dependent samples), bleaching/contamination during measurements leading to a reduction in CL signal, fluctuations in current and mirror alignment.\\ \par

{\bf Supplementary information} The Supporting Information provides further information about the calculation of the Mueller matrix, the calibration of set up, the spectral response of the antennas and the data analysis for the bulk material measurements. \\ \par


{\bf Notes} A.P. is co-founder and co-owner of Delmic BV, a startup company that develops a commercial product based on the ARCIS cathodoluminescence system that was used in this work. \\ \par

{\bf Acknowledgements} The authors thank Abbas Mohtashami for his help with the fabrication and Henk-Jan Boluijt for the cartoon in Fig.~\ref{Fig1}(a). This work is part of the research program of the Stichting voor Fundamenteel Onderzoek der Materie (FOM), which is financially supported by the Nederlandse Organisatie voor Wetenschappelijk Onderzoek (NWO). This work is part of NanoNextNL, a nanotechnology program funded by the Dutch ministry of economic affairs. It is also supported by the European Research Council (ERC).

\section{Supporting Information}

\subsection{Spectral measurements on bullseye}

\begin{figure*}[thb!]
\centering
\includegraphics[width=0.8\textwidth]{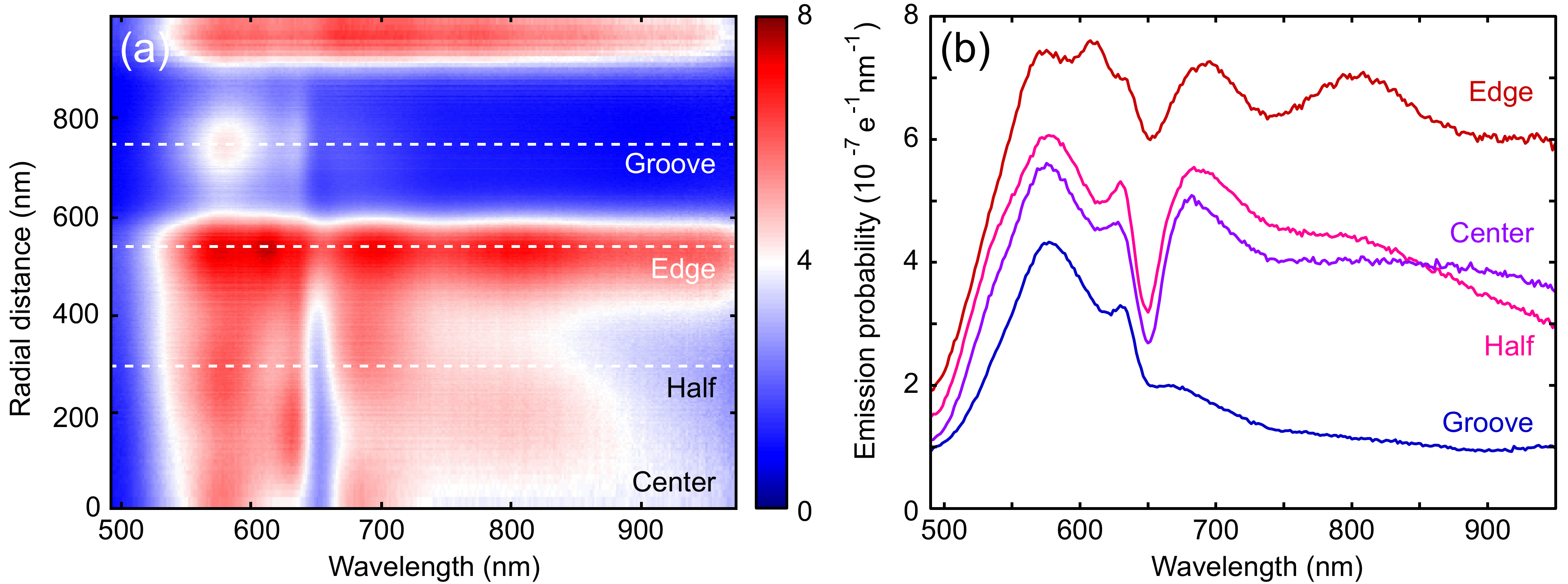}
\caption{(a) CL intensity as a function of wavelength and radial position from the center of the bullseye. The spectra are corrected for dark noise and the spectral response of the detection system. (b) CL spectra at selected radial positions corresponding to the center, halfway, and the edge of the bullseye plateau. We also show a spectrum for excitation in the center of the first groove. The positions are indicated by the white dashed lines in (a). Notice that the color bar in (a) and the vertical axis in (b) both correspond to emission probability.}
\label{FigS1}
\end{figure*}

 To study the spectral response of the bullseye from Fig. 2 (main text), we raster scan the electron beam over the central part of the structure in 15 nm steps, and collect a CL spectrum for every pixel using a visible/NIR fiber-coupled Czerny-Turner spectrometer~\cite{coenen_NL11}. The scan includes the central plateau, the first groove, and part of the first ridge. The spectra are corrected for the system response using the TR emission of the unstructured gold substrate~\cite{kuttge_PRB09}. We then radially average the spectrum to obtain a map showing emitted intensity as a function of wavelength and radius, taking into account the proper Jacobian so we can directly compare the intensities (shown in Fig.~\ref{FigS1}(a)). Figure \ref{FigS1}(b) shows individual spectra for specific radial distances as indicated in (a) by the white dashed lines.

We observe several features in the scan. On the plateau, the emission is quite broadband except for a strong dip in intensity around $\lambda_{0}$ = $650$ nm. The CL intensity increases towards the edge of the plateau, then drops in the groove, and becomes bright on the first ridge again. The observed spectral features can be attributed to a mix of resonant and diffractive effects.

 Because the groove is wide enough, it supports a zero-th order standing wave mode on the bottom of the groove around $\lambda_{0}$ = 580 nm, which is cut-off for longer wavelengths leading to a low intensity in the red~\cite{Brucoli_PRB11,Schoen_NL13}. The dip at $\lambda_{0} = 650$ nm can be attributed to an interesting experimental artifact related to the diffraction of SPPs. Consistent with the grating equation mentioned in the main text, the bullseye emits very close to the normal at this wavelength. Because the bullseye structure is highly directional, a major part of this diffracted beam is lost through the $600$~\textmu m hole in the mirror right above the sample. In fact, one could use this dip to find the wavelength and spatial position at which such a structure attains maximum directionality in the normal direction. In this case, the effect is strongest for the central positions in the bullseye because the beam is exactly normal to the sample  as is visible in Fig. 3(a) for $\lambda_{0}$ = $750$ nm. Even though the dip is strongest in the center of the plateau, it is visible for every radial excitation position in this map (even within the groove), indicating that the extended bullseye geometry always causes a significant fraction of CL emission to be in the normal direction.

\subsection{Calculation of the Mueller matrices}

The Mueller matrix of an optical element accounts for the effect of the element on the polarization state of an incident field \cite{Born_Wolf}. The Mueller matrices of a linear polarizer and a QWP are well-known for example. We use the Mueller formalism to relate the Stokes vector in the detection plane to the Stokes vector describing the sample emission polarization. The resulting Mueller matrix of the mirror contains both geometric and polarizing effects of the mirror in the emission polarization.

To retrieve the Mueller matrix of our light collection system, we calculate how the electric field components $E_{\theta}$ and $E_{\varphi}$ in the emission plane transform to $E_{y}$ and $E_{z}$ in the detection plane~\cite{Bruce_OPT06,Bruce_04}. To that end we calculate how fully isotropic $p$-polarized and $s$-polarized emission is projected onto the detection plane by the parabolic mirror. We use the geometrical methods described in section 3 of Ref.~\cite{coenen_OE12} to account for the parabolic reflector and we use the full complex Fresnel reflection coefficients to accurately describe the light reflection on the mirror. These coefficients were calculated using tabulated optical constants~\cite{Palik} for the central frequency of the collection bandwidth. As the reflection angle at the mirror is different for every wave-vector emanating from the sample, each element of the Mueller matrix is a function of the emission angle \cite{Bruce_OPT06,Bruce_04}. Instead of calculating the Mueller matrix, one could also envision experimentally retrieving it. This requires a precisely controlled radial and azimuthal polarization source as standard. While transition radiation could serve as a fully radial source, a fully azimuthal source is not readily available.

The Mueller matrices can be used in two directions. Either one can invert measured data from the detector plane to sample coordinates, or in the opposite direction, one can predict how a given source will appear on the detector plane. For our analysis of TR emission, in Fig.~\ref{FigS2} we apply Mueller matrices to theoretical TR emission  to predict the measured data for each setting of the polarimeter. In this case, we combine the mirror Mueller matrix with the Mueller matrices of a linear polarizer and a QWP, which are a function of the selected analyzer angles $\alpha$ and $\beta$ \cite{Chipman}. We note that for fully polarized sources this is analogous to the approach in Ref.~\cite{coenen_OE12} where the Jones matrix of the polarizing element operates on the Jones electric field vector.

\subsection{Polarimetry of transition radiation emission}

\begin{figure*}[htb!]
\centering
\includegraphics[width=0.7\textwidth]{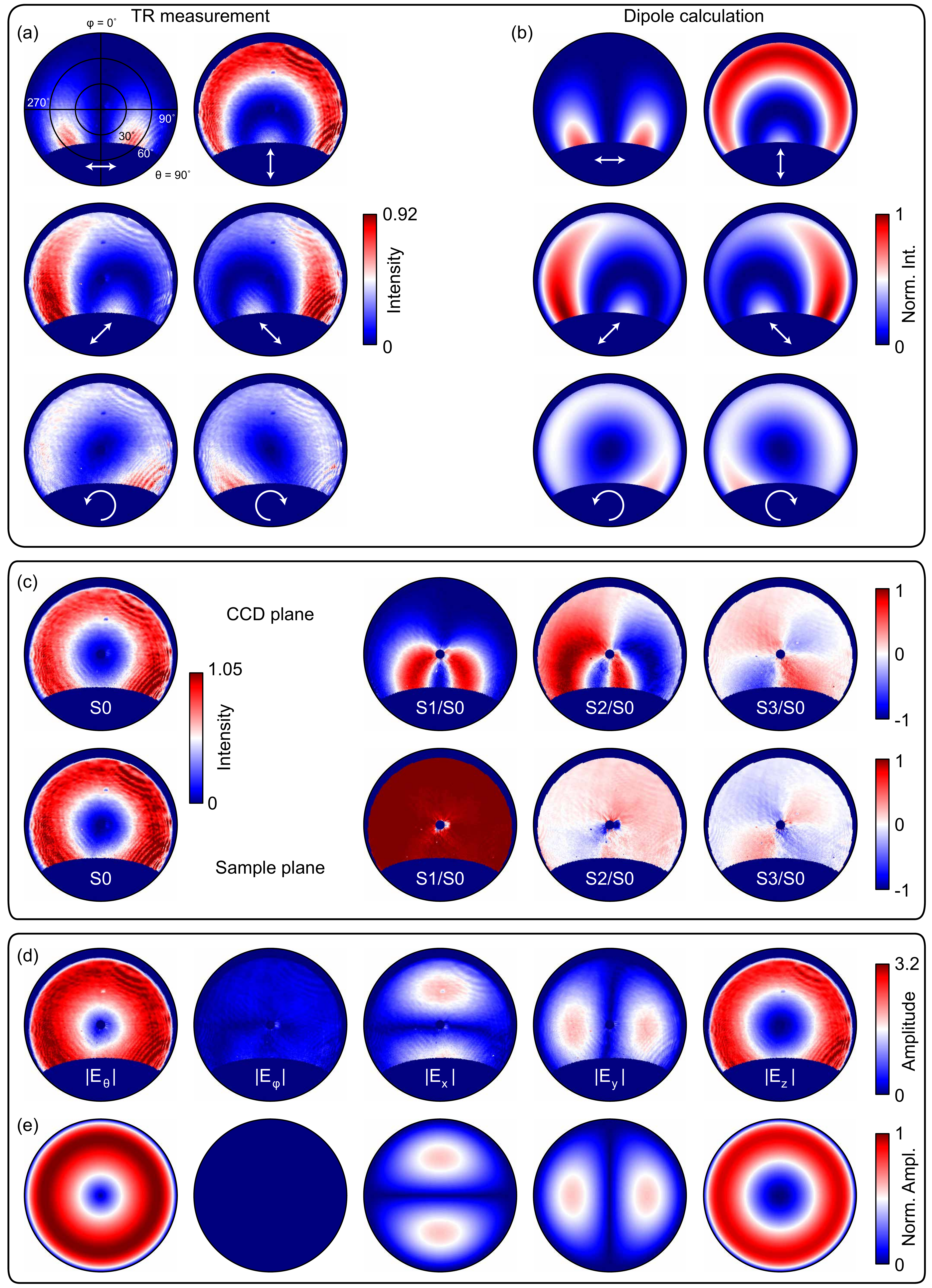}
\caption{(a) Polarization-filtered angular CL patterns of TR emission for a Au single crystal for different analyzer settings as indicated by the white arrows, measured at $\lambda_{0} = 850$ nm. (b) Calculated polarization-filtered patterns for a vertically oriented point-dipole source on top of a gold substrate (at $\lambda_{0} = 850$ nm). (c) Stokes parameters as a function of angle in the CCD plane as well as in the sample plane, after the mirror correction. The $S_{1}$, $S_{2}$, and $S_{3}$ patterns are normalized to $S_{0}$ to better show the overall polarization distribution. (d) Spherical and cartesian field amplitude distributions as a function of angle, retrieved from the experimental data in (a). (e) Calculated field amplitudes for a vertical dipole.}
\label{FigS2}
\end{figure*}

Transition radiation (TR) emission occurs whenever an electron traverses an interface between two dielectric media. The electron locally polarizes the material close to the interface, giving rise to a well-defined broadband vertically-oriented point-dipole-like source~\cite{revjav,coenen_APL11,coenen_OE12}. This makes it a useful source to test our CL polarimetry technique. Here we perform polarimetry measurements on an unstructured part of the single-crystal Au substrate from which only TR emission is expected, using the same technique used for the bullseye. Figure \ref{FigS2}(a) shows the TR data at $\lambda_{0}$ = $850$ nm for every polarimeter setting. The top left panel includes the angular coordinate system for reference.

In addition to the TR measurements, we calculated the emission pattern for a z-oriented dipole on top of an Au substrate and its polarization components. The dipolar far-field for $\lambda_{0} = 850$ nm was calculated from the asymptotic far-field expressions~\cite{nanooptics} using ellipsometry data for the optical constants of the gold substrate. We then calculate the expected filtered pattern using the appropriate Mueller matrices for the paraboloid mirror and the polarimeter components. Figure~\ref{FigS2}(b) shows the result of this calculation including the angular acceptance of the mirror, in order to allow a good comparison with the data. The excellent agreement of both angular distributions and relative intensities between data and calculation indicates that TR emission is indeed dipolar and that the mirror correction works well. The fringes in the data are an experimental artifact, probably due to interference between multiple reflections of the optical elements on the detector.

We can use the measurements from Fig.~\ref{FigS2}(a) to determine the Stokes parameters in the detection plane of the CCD and then use the Mueller matrix formalism to correct for the effects of the mirror. This allows us to retrieve the Stokes parameters in the sample plane from which we can determine the different field amplitudes. Figure~\ref{FigS2}(c) shows the Stokes parameters in both the detection and sample planes, where $S_{1}$, $S_{2}$ and $S_{3}$ have been normalized by $S_{0}$ so that we can clearly observe the polarization distortions due to the mirror. In the detection plane the Stokes parameters display complex patterns that are very similar to those shown for the bullseye in Fig. 2(c), since both cases are dominated by purely radial polarization. In the sample plane the behavior of the Stokes parameters is much simpler, $S_{0}$ has barely changed and $S_{1}$ is close to 1 while $S_{2}$ and $S_{3}$ are very small. Again, TR is expected to be fully radially polarized, so there should be no diagonal ($S_{2}$) or circular ($S_{3}$) components. The striking difference between the Stokes parameters in the two planes underscores the importance of the Mueller matrix correction to provide accurate results.

Fig \ref{FigS2}(d) shows the spherical and Cartesian field amplitudes that have been retrieved from the Stokes parameters in sample space. The fields nicely reveal the expected radially polarized nature of the emission. The amplitudes have been plotted using a single color scale to allow a quantitative comparison between the different components. Both the relative amplitudes and amplitude distributions match very well with the calculated dipolar fields shown in (e). These results demonstrate that cathodoluminescence polarimetry can reliably be used as a quantitative tool for deducing the far-field polarization distribution of a nanoscale emitter.

\subsection{Silicon and GaAs polarimetry}

\begin{figure*}[thb]
\centering
\includegraphics[width=0.9\columnwidth]{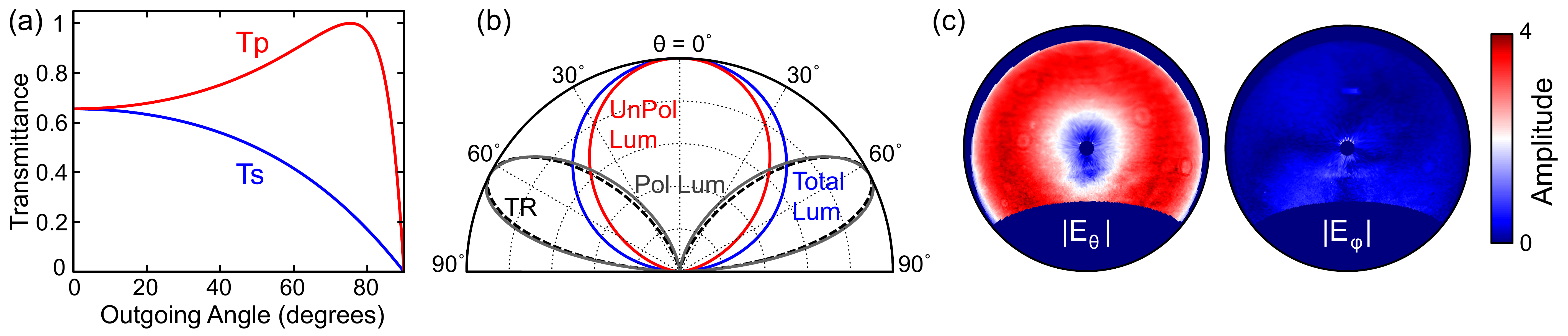}
\caption{(a) Transmission coefficients $T_{p}$ and $T_{s}$ at the sample-vacuum interface at $\lambda_{0}$ = 650 nm for Si, as a function of the internal angle of emission. (b) Calculated normalized angular emissions patterns of the different contributions to Si CL emission. The total luminescence profile (blue) as well as the unpolarized (red) and polarized (grey) luminescence contributions are shown together with the theoretical TR profile (black dashed line). (c) $|E_{\theta}|$ and $|E_{\varphi}|$ field amplitudes of the polarized emission from Si, in units of $10^2 \sqrt{ADU sr^{-1} s^{-1}}$.}
\label{FigS4}
\end{figure*}

Calculating the contributions of TR, polarized and unpolarized luminescence to the total emission from Si or GaAs requires determining their angular profiles. An essential part of this process are the transmission coefficients $T_{p}$ and $T_{s}$ at the sample-vacuum interface, shown in Fig. 5(a) for the case of Si at $\lambda_{0}$ = 650 nm. The large contrast between the two coefficients at angles above $\sim$ 20$^{\circ}$ leads to more $p$-polarized light exiting the Si than $s$-polarized light. Especially near the Brewster angle, the $p$-component of the field is transmitted significantly better than the $s$-component, with a contrast between intensity transmission $T_{p}$ and $T_{s}$ exceeding 4 (see Fig. 5(a)). This difference in transmission coefficients results in a CL emission profile slightly different than the Lambertian cos($\theta$) profile (blue curve in Fig.~\ref{FigS4}(b)). The polarizing effect of the interface does not noticeably affect the emission pattern of the total luminescence, but the unpolarized luminescence (consisting of equal amounts of $s$- and $p$-polarized light) is markedly narrower. Accordingly, the polarized part of the luminescence is stronger at higher angles and, interestingly, it follows a very similar profile to that of TR for Si at $\lambda_{0}$ = 650 nm, which has been calculated using formulas derived in Ref.~\cite{revjav}. The same formulas are used to calculate the gray line describing TR from Au at $\lambda_{0}$ = 850 nm in Fig. 5(a).

Once the theoretical profiles for the different emission processes are calculated, it is possible to determine their relative contributions to the total emission. The fraction of polarized and unpolarized luminescence is fully specified by the Fresnel equations. This gives enough information to compare calculations to data from GaAs with very good agreement, as that is fully dominated by luminescence. The case of Si is more complex as TR also plays a role, so the polarized emission is comprised of coherent TR as well as polarized luminescence. We determine the ratio of TR and luminescence by fitting the total intensity ($S_{0}$) to a linear combination of the two processes. Fresnel calculations again predict the polarized and unpolarized contributions to the luminescence so that we can combine all three components. For both the calculations and the experiments we scale the angular distributions by the overall (integrated) emission intensity, and find good (absolute) agreement, as was shown in Fig. 5(b). The polarized signal constitutes $\sim 32\%$ of the total emission. More explicitly, we find that TR contributes $\sim 21\%$ of the total CL intensity, so polarized luminescence contributes $\sim 11\%$ and unpolarized luminescence $\sim 68\%$. 

After separating the polarized and unpolarized components we can retrieve the different electric field components. Both the TR and the polarized luminescence should be $p$-polarized, which we verify from the experimental data using the Mueller analysis to determine the radial and azimuthal field amplitudes. This is shown for Si in Fig.~\ref{FigS4}(c), where we indeed observe that almost all of the amplitude is in the $|E_{\theta}|$ component, i.e. for $p$- polarization. This demonstrates that we can separate the unpolarized and polarized emission even from mostly incoherently radiating semiconductors and still retrieve the correct electric fields for the polarized portion of the emission.

\bibliography{references_polarimetry}

\end{document}